\documentclass[referee]{raa}
\usepackage[hyperindex,pdftex]{hyperref}
\usepackage{graphicx, times}             
\usepackage{natbib}
\usepackage{amssymb, amsmath}
\bibpunct{(}{)}{;}{a}{}{,}
\hypersetup{colorlinks = true, linkcolor = green, anchorcolor = red, citecolor = blue, filecolor = red, pagecolor = red, urlcolor = red}

\begin{document}
\title{Synthesising Solar Radio Images From Atmospheric Imaging Assembly Extreme-Ultraviolet Data}

\volnopage{Vol.0 (20xx) No.0, 000--000}      
\setcounter{page}{1}          

\author{Z. F. Li, S. H. Hua, X. Cheng \& M. D. Ding}
\institute{School of Astronomy and Space Science, Nanjing University, Nanjing 210023, China; {\it zhuofeili98@gmail.com; xincheng@nju.edu.cn}\\
}

\abstract{During non-flaring times, the radio flux of the Sun at the wavelength of a few centimeters to several tens of centimeters mostly originates from the thermal bremsstrahlung emission, very similar to the EUV radiation. Owing to such a proximity, it is feasible to investigate the relationship between the EUV emission and radio emission in a quantitative way. In this paper, we reconstruct the radio images of the Sun through the differential emission measure obtained from the multi-wavelength EUV images of the Atmospheric Imaging Assembly on board \textit{Solar Dynamic Observatory}. Through comparing the synthetic radio images at 6 GHz with those observed by \textit{Siberian Radioheliograph}, we find that the predicted radio flux is qualitatively consistent with the observed value, confirming thermal origin of the coronal radio emission during non-flaring times. The results further show that the predicted radio flux is closer to the observations in the case of including the contribution of the plasma with temperatures above 3 MK than in the case of only involving the low temperature plasma as was usually done in the era of pre-\textit{SDO}. We also discuss the applications of the method and uncertainties of the results.}

\maketitle

\section{Introduction}        
\label{sect:intro}
Solar extreme-ultraviolet (EUV) radiation refers to the electromagnetic radiation at wavelengths  of 100--1200 {\AA}. It is of thermal origin and generated mainly through line emission of highly ionized ions, whose intensity depends on the plasma temperature and density, or more precisely the emission measure (EM) \citep{zhang2001reconciling}. On the other hand, solar radio emission usually includes incoherent emission and coherent emission. Thermal bremsstrahlung and gyroresonance emission of free electrons are incoherent emission, while the coherent emission is resulting from nonthermal electrons, including electron cyclotron maser emission and plasma emission, etc. \citep{kundu1965solar}. Bremsstrahlung radiation originates from free-free emission through Coulomb collisions between electrons and ions \citep{wild1963solar}. The emission flux from bremsstrahlung is also dependent on the plasma EM, similarly to the EUV emission. By contrast, gyroresonance emission is generated by non-relativistic electrons in magnetic fields, whose flux is thus not only in dependence on the EM but also on the magnetic field. For both of the EUV and radio radiations, their temporal evolution generally includes two components: a slowly varying component (the background) from the quiet Sun and an impulsively varying one from flares in active regions \citep{kundu1965solar}. For active regions, in non-flaring periods, the EUV flux is almost dominated by thermal bremsstrahlung emission, while the radio flux is also mostly from bremsstrahlung with some possible contribution from non-thermal gyroresonance emission. However, for quiescent regions, EUV and radio emissions are almost fully contributed by bremsstrahlung as the magnetic field over there is very weak so that the gyroresonance emission can be ignored \citep{shibasaki2011solarphy}. 

Because of the similar origin for EUV and radio emissions in non-flaring times, a quantitative comparison between them can help diagnose the physical properties of the corona. Using the data from the Extreme-Ultraviolet Imaging Telescope (EIT) and assuming a corona with two temperature components in the range of 0.5--3 MK, \cite{zhang2001reconciling} determined the EM of each pixel over the full disk for the hot and cool components. They further calculated the brightness temperature ($T_b$) of radio emission using the formula:
\begin{equation}
T_b=0.2\nu^{-2}T_{C}^{-0.5}EM_{C}+0.2\nu^{-2}T_{H}^{-0.5}EM_{H}
\label{eq:tb}
\end{equation}
where, $\nu$ denotes the frequency in the radio domain, $T_{C}$ ($T_{H}$) and $EM_{C}$ ($EM_{H}$) are the temperature and EM of the cool (hot) plasma component, respectively. Note that, the formula is valid only if the emission is optically thin, i.e., the optical depth ($\tau$) is much smaller than 1. As a result, they found that the predicted and observed VLA (Very Large Array) radio images show very similar morphologies and there exists a good linear correlation between the predicted brightness temperature and the observed one. However, they also noted that the predicted flux is systematically larger than the observed one, which was attributed to an underestimation of the abundance of iron relative to hydrogen when calibrating the EIT data.

Nevertheless, with the launch of \textit{Solar Dynamics Observatory} \citep[\textit{SDO};][]{Pesnell2012}, it is realised that the coronal plasma is actually distributed in a very wide temperature range. In particular, for flaring active regions, the temperature range could be 0.5--20 MK \citep[e.g.,][]{Cheng2012,Hannah2012,Cheung2015}. Even during non-flaring times, the emission from the plasma with temperatures of above 3 MK is still nontrivial \citep[e.g.,][]{Hannah2016,Grefenstette2016}. This means that the hot plasma could also have a considerable contribution to the radio emission, especially in radio wavelength range of centimeters to several tens of centimeters. It is thus more reasonable to include the entire thermal plasma in the corona to predict the radio emission. In this paper, using the data from the Atmospheric Imaging Assembly \citep[AIA;][]{Lemen2012}, we calculate the differential emission measure (DEM) for each pixel over the full disk, with which a map of $T_b$ is estimated under the assumption of an optically thin corona. The main purpose is to revisit the relationship between the predicted radio emission and the observed one during non-flaring times. The observational data and analytical procedure are introduced in Section \ref{sect:data}, which is followed by the results in Section \ref{sect:results}. A summary and discussions are given in Section \ref{sect:conclusion}.

\section{Data selection and analyses}
\label{sect:data}
The AIA on board $\textit{SDO}$ observes the Sun's atmosphere in seven EUV channels, six of which, centered at 94~\AA\,, 131~\AA\,, 171~\AA\,, 193~\AA\,, 211~\AA\, and 335~\AA\,, are used for DEM analysis. Each AIA EUV image contains 4096 by 4096 pixels with a spatial resolution of 1.2$''$, covering a field of view of 1.3 R$_\odot$. The time cadence of the images is 12 s. These data can be freely downloaded from the $\textit{SDO}$ website\footnote{$http://jsoc.stanford.edu/ajax/exportdata.html$}.

The full disk radio images are provided with \textit{Siberian Radioheliograph} (\textit{SRH}), which was constructed as an upgrade of the \textit{Siberian Solar Radio Telescope} (\textit{SSRT}). It observes the Sun at 5.7 GHz with a routine mode but switched to a mode for high-cadence and high-resolution observation when solar radio bursts were detected \citep{Grechnev2003}. SRH started routine observations in August of 2016 at several frequencies in the range of 4--8 GHz with an angular resolution of 1--2$'$ and a cadence of about 12 s \citep{Lesovoi2012}. Here, we use the full disk radio images at 6.0 GHz observed on 2016 March 16  for analysis.

\begin{figure}[!ht]
\centering
\includegraphics[width=15cm]{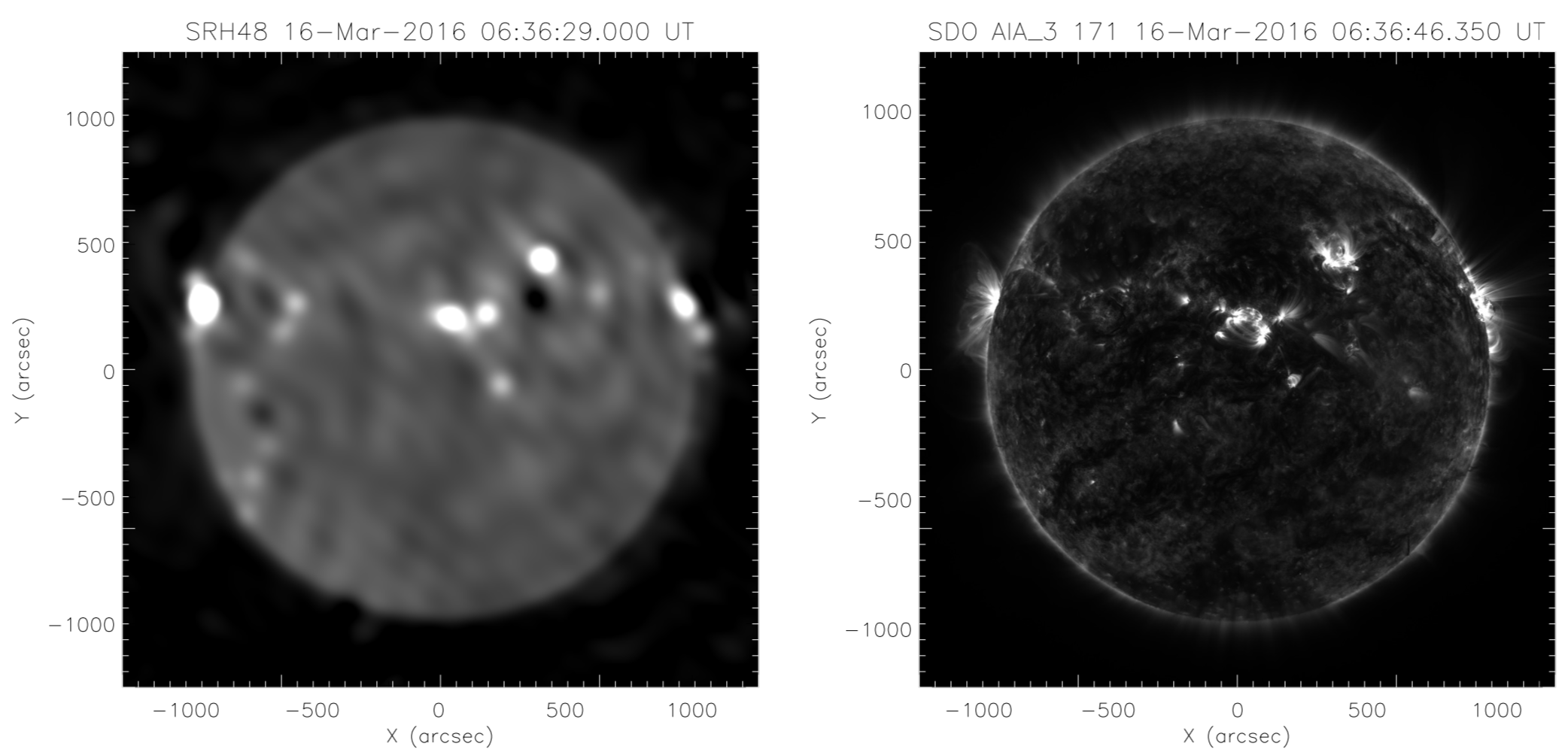}
\caption{Left: Full disk 6 GHz radio image on 2016 March 16 observed by \textit{SRH}. Right: Corresponding full disk EUV image observed by AIA at the 171 \AA\ passband at a closest time.}
\label{msfig1}
\end{figure}

We also make use of the 17 GHz radio images that were observed by \textit{Nobeyama Radioheliograph} (\textit{NoRH}). NoRH provides one full disk image every one second at two frequencies of 17 and 34 GHz since June of 1992. The spatial resolution is about $10''$ at 17 GHz and $5''$ at 34 GHz. The images are reconstructed through the CLEAN algorithm \citep{HNakajima1994}. The 17 GHz full disk image on 2016 March 16 is downloaded from the \textit{NoRH} homepage\footnote{$https://solar.nro.nao.ac.jp/norh/$}.

In order to predict the parameter $T_b$ at each pixel over the full disk, we need to first determine the  DEM at this pixel as a function of temperature. Theoretically, for an optically thin medium, the emission measure (EM) is defined as
\begin{equation}
EM=\int~n_e(s)n_H(s)ds
\label{eq:EM}
\end{equation}
where $n_e$ is the electron number density, $n_H$ is the hydrogen number density, and \emph{s} is the depth along the line of sight. The DEM is 

\begin{equation}
  DEM(T)=\frac{dEM}{dT}=\int\frac{dn_e(s)n_H(s)}{dT(s)}ds
\label{eq:DEM}
\end{equation} 
The EUV intensity at a specific passband $i$ can be denoted as
\begin{equation}
  I_i=\int R_i(T)DEM(T)dT
\label{eq:intensity}
\end{equation}
where $R_i(T)$ is the temperature response function of the AIA detector at this passband. With the observed EUV images at the six passbands, the DEM at each pixel can be derived through resolving an inverse problem even though ill posed. The nonlinear response functions make the inversion very complicated. Here, we make use of the code developed by \cite{cheung2015thermal}, which uses a matrix inversion to ensure non-negative solutions. The code seeks a solution with an optimal error resolution and the computation is faster than previous DEM inversion methods. Thus, it is very appropriate for the present project. Note that in this work, we do not need to know the exact values of $n_{e}$ and $n_{H}$, but just deduce the DEM value in order to quantify the brightness temperature as shown below.
 
After knowing DEM, the brightness temperature can be predicted at a given frequency. The optical depth $\tau$ of thermal bremsstrahlung is proportional to the EM and is also a function of the electron temperature $T_e$ and the frequency $\nu$. At radio wavelengths, where the Rayleigh-Jeans limit is valid, it is reasonable to quantify radio flux with $T_b$ (\citep{zhang2001reconciling}), which can then be expressed as
\begin{equation}
T_b=\int _{T_{min}} ^{T_{max}}  0.2 \nu ^{-2}T^{-1/2}DEM(T) dT.
\label{eq:T_b}
\end{equation}
In the current work, the temperature is integrated over the range of 0.5--10 MK, where the DEM solutions have been tested and are relatively reliable \citep[e.g.,][]{Cheng2012}.

\begin{figure}[!h]
\centering
\includegraphics[width=15cm]{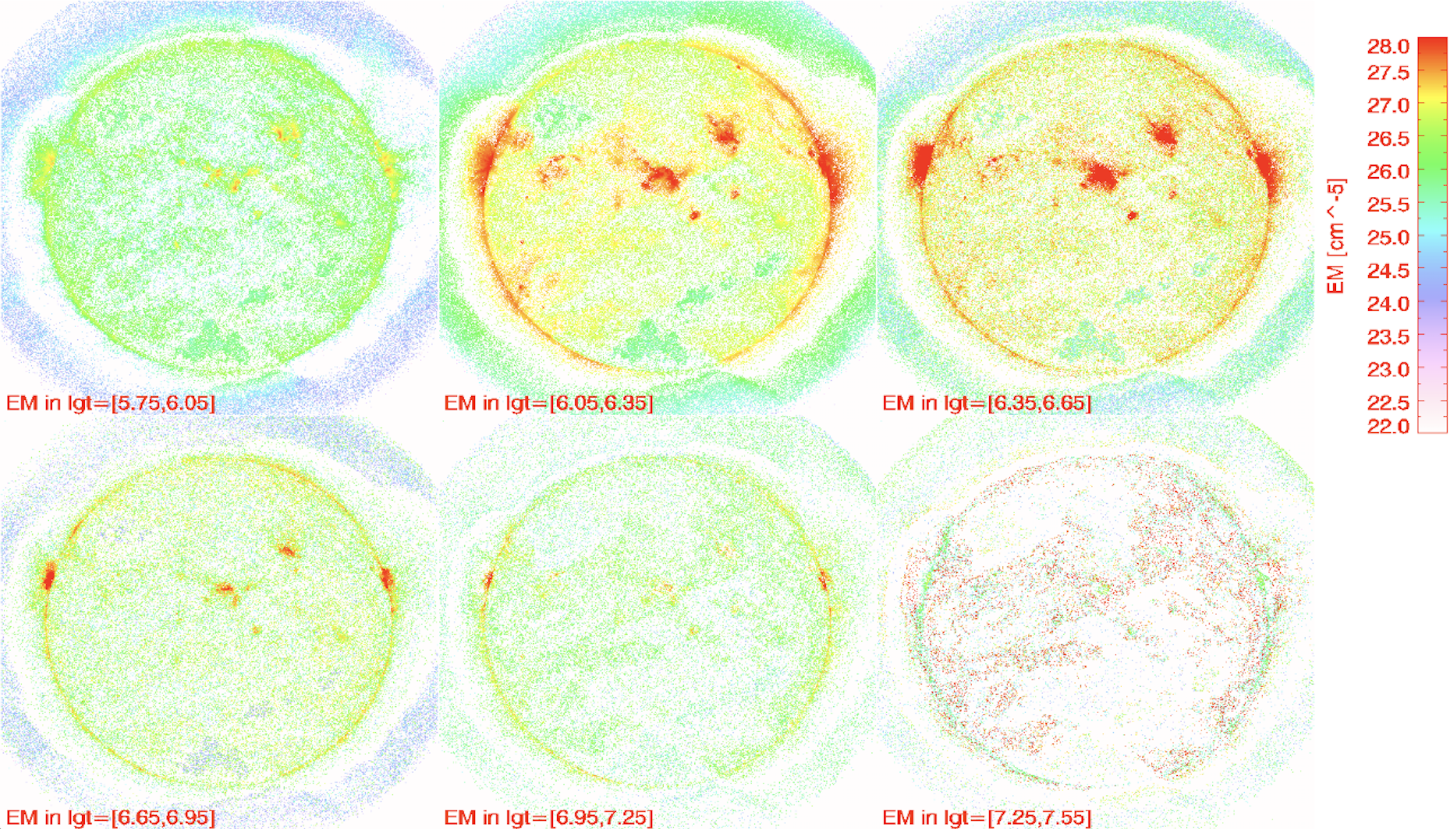}
\caption{EM maps at different temperature intervals.}
\label{msfig2}
\end{figure}

\section{Results}
\label{sect:results}
The full disk solar radio image at 6 GHz on 2016 March 16 observed by SRH is shown in the left panel of Figure~\ref{msfig1}. For comparison, the AIA EUV image at the 171 \AA\ passband at the same time is shown in the right panel of Figure~\ref{msfig1}. One can see that the locations with strong radio emission correspond to the active regions that can be clearly identified with enhanced 171 \AA\ emission very well. However, the radio image is much more blurry than the 171 \AA\ image, which is mainly due to the fact that the latter has a much higher spatial resolution ($1.2''$) than the former ($9.6''$). 

To reconstruct the radio images from the EUV data, we first calculate the DEM at each pixel of the full disk using six AIA EUV passbands. Here, the AIA images are degraded to the same resolution and can thus be directly compared with the radio data. The temperature range is set as 5.7$\le$ $\log T$ $\le$7.7 with an interval of $\Delta \log T$=0.1, corresponding to 0.5--50 MK.

In Figure~\ref{msfig2}, we show the EM maps of the full disk at the different temperature intervals. We can see that the emission in active regions is primarily from the plasma in the temperature range of 1-5 MK. The EM map at the temperature interval of 7.2$\le$ $\log T$ $\le$7.5 shows a lot of high EM points (in red) which turn out to be incorrect after a careful inspection of the corresponding DEM curves. Thus, we restrict the temperature range to be less than $\log T=7.0$ when calculating the EM and then $T_b$ following the equation \ref{eq:T_b}. In order to compare our results with that of \cite{zhang2001reconciling}, we also calculate the parameter $T_b$ using the two-temperature model they proposed, in which the cool component has a temperature range from 0.8 MK to 1.4 MK and the hot one is from 1.6 MK to 2.8 MK. In the following, we also make the integration for temperature ranges of 0.5--10 MK called full-temperature model. The comparison between the two models can help reveal how much contribution is from the plasma above the temperature of 3 MK to the thermal radio emission.

\begin{figure}[!h]
\centering
\includegraphics[width=15cm]{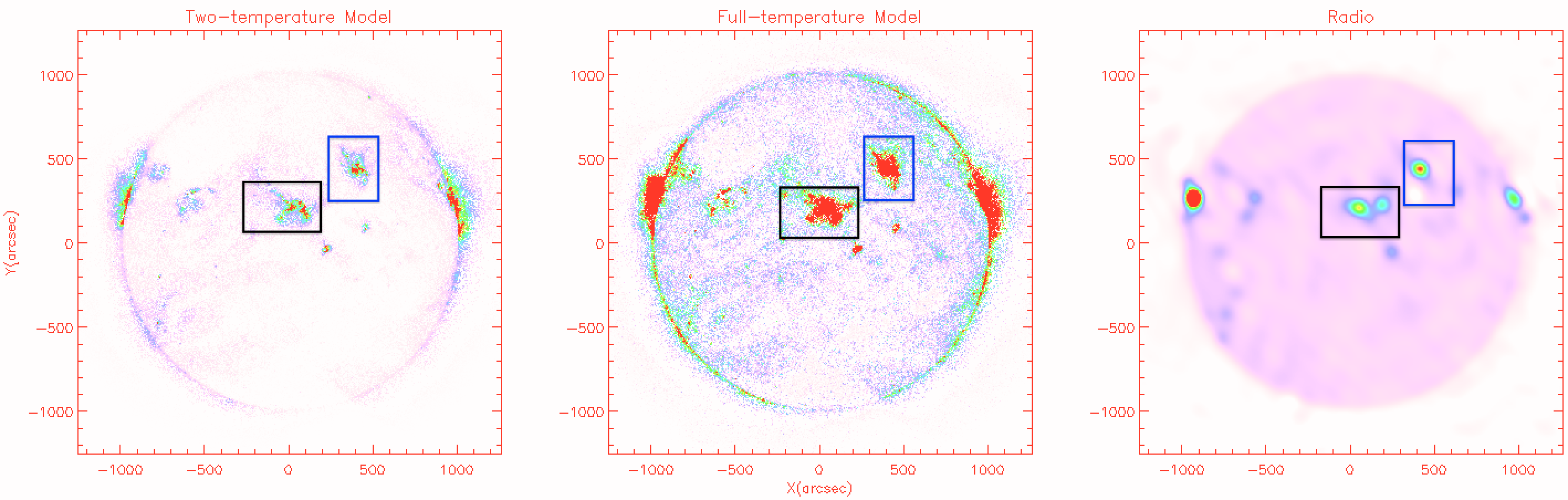}
\caption{Left: Predicted $T_b$ image of the full solar disk at 6 GHz calculated by the two-temperature model. Middle: Predicted $T_b$ image at 6 GHz from the full-temperature model. Right: Observed radio image at 6 GHz from \textit{SRH}. The two boxes indicate the active region 12519 (black) and 12521 (blue).}
   \label{msfig3}
   \end{figure}
   
\begin{figure}[!h]
\centering
\includegraphics[width=15cm]{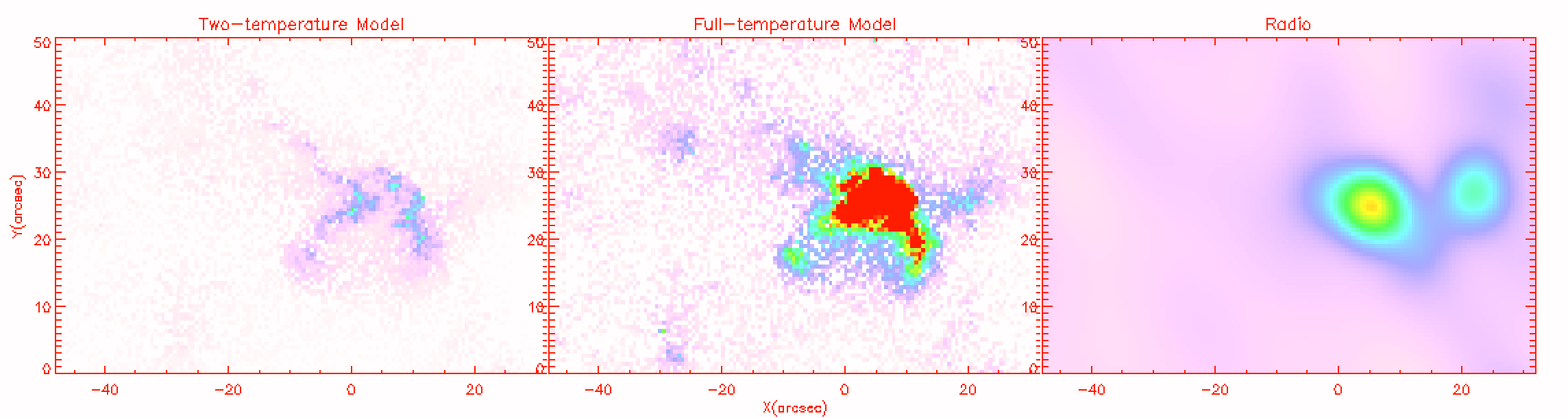}
\caption{Same as Figure \ref{msfig3} but for the active region 12519 located at the disk center.}
\label{msfig4}
\end{figure}

Through equation~\ref{eq:tb} and \ref{eq:T_b}, we reconstruct the $T_b$ image of the full disk at 6 GHz as shown in Figure~\ref{msfig3}. One can see that the locations of active regions in the reconstructed images are basically consistent to that in the observed images. However, the predicted images from the full-temperature model seem to show more structures and stronger emission than the images from the two-temperature model, which can be more apparently revealed in the selected active region 12519 as shown in Figure~\ref{msfig4} (also marked with black box in Figure~\ref{msfig3}). This active region just appears as an oval shape in the observed radio image but shows two branches extending toward the south in the predicted image. With the same contrast, the full-temperature model shows the structures in active regions more clearly and displays a $T_b$ distribution at the active region center more consistent to observations. It suggests that the full-temperature model is a better one to estimate the radio emission from EUV observations.
 
\begin{figure}[!ht]
\centering
\includegraphics[width=15cm]{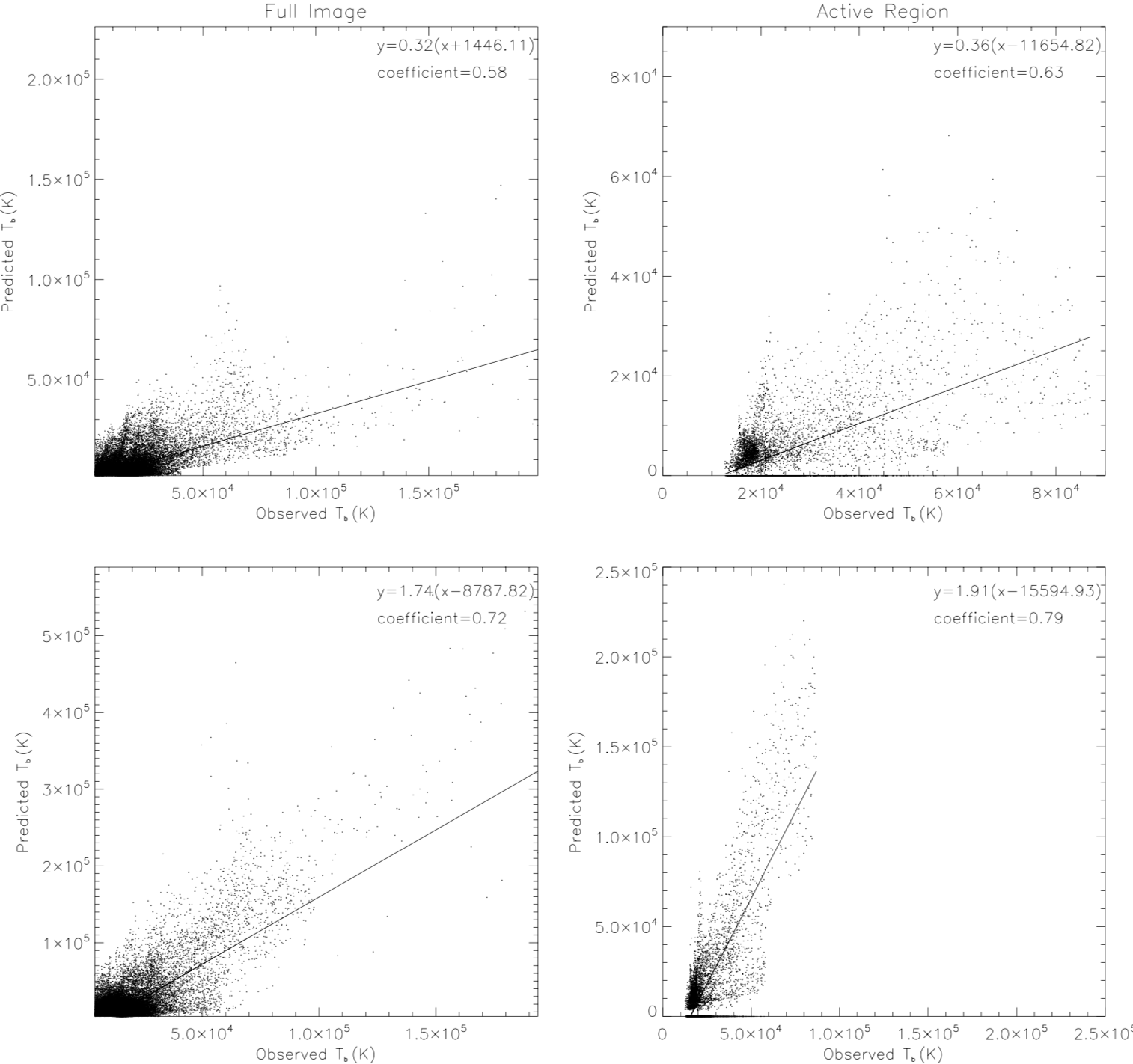}
\caption{\small Pixel-by-pixel correlation plots between the predicted $T_b$ at 6 GHz from the two-temperature model and the observed one for the full disk (upper left) and a selected active region (upper right). The solid lines denote the linear fitting to the data points. The fitting formulas and correlation coefficients are also indicated in each panel. Lower: The correlation results between the full-temperature model and the observed value are also displayed for the full disk (lower left) and the active region (lower right), respectively.}
\label{msfig5}
\end{figure}

\begin{figure}[!ht]
\centering
\includegraphics[width=15cm]{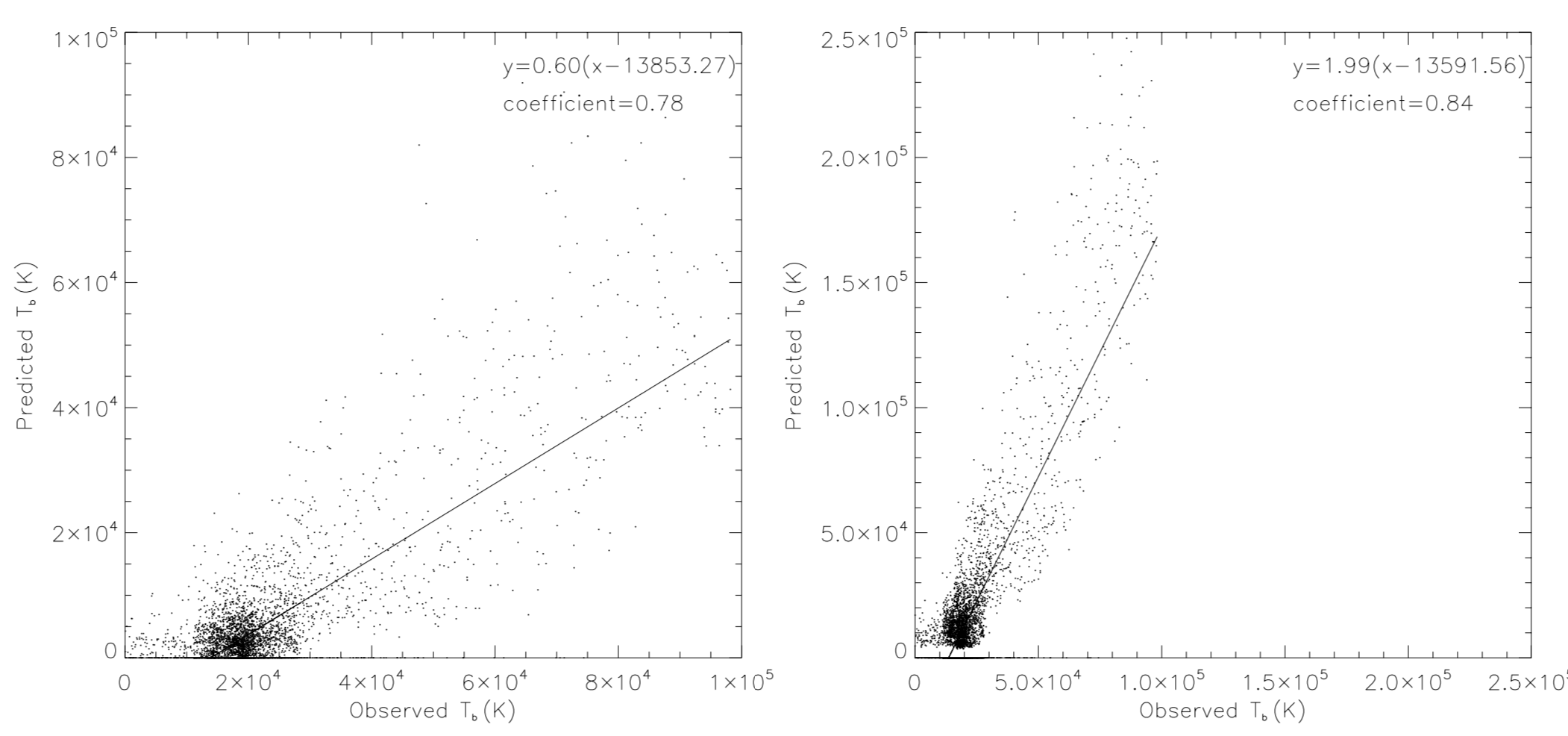}
\caption{Pixel-by-pixel plots for the active region 12521 showing a good linear correlation between the predicted and observed $T_b$. Left: The correlation results between the two-temperature model and the observed value. Right: The correlation results between the full-temperature model and the observed value.}
\label{msfig6}
\end{figure}

For the purpose of quantitatively investigating the relationship between the predicted $T_b$ and observed one, we make scattering plots, together with linear fitting to them, for all the pixels in the full disk and the selected active region, respectively. The $T_b$ values for most pixels both from the predicted and the observed images are very small, even close to zero,  which were thought to be unreasonable, and thus are removed. In the two-temperature model, pixels whose values are less than 2000K are removed, and in the full-temperature model, the boundary is 4000K. The results are presented in Figure~\ref{msfig5}. One can see that the two-temperature model predicts the $T_b$ value at 6 GHz roughly in a linear relationship to the observed one, for both the full disk (upper left panel of Figure~\ref{msfig5}) or a specific active region (upper right panel of Figure~\ref{msfig5}). The linear correlation coefficients are found to be 0.58 and 0.63 for the full disk and the active region, respectively. Moreover, it is found that for the full disk the predicted $T_b$ is mostly distributed in the range of 1000 K--0.1 MK, systematically smaller than the observed one as revealed by the slopes of the fitting lines, 0.32 and 0.36. Interestingly, when inspecting the results from the full-temperature model, we find that the correlations between the predicted $T_b$ and observed one become better for both the full disk and the active region (lower left and lower right panels of Figure~\ref{msfig5}). The corresponding linear correlation coefficients are 0.72 and 0.79, respectively. However, in this case the predicted $T_b$ is systematically larger than the observed one. The slopes of the fitting lines are 1.74 and 1.91 for the full disk and the active region, respectively. 
Note that the best fitting requires non-zero constants to be added to or subtracted from the observed value of $T_{b}$. These constants in the fitting functions may result from the deconvolution process of making radio images.
The above results show that the full-temperature model may be more accurate than the two-temperature one in predicting the radio emission of thermal origin and that the radio emission at low frequencies is mostly dominated by emission from the thermal plasma.

We also make the same analysis to the active region 12521 (marked with blue box in Figure~\ref{msfig3}) and get a good linear correlation between the predicted and observed $T_b$ (Figure~\ref{msfig6}). Besides, we analyze the full disk without any active regions, and find that the correlation becomes very bad. So we conclude that during non-flaring period, the radio emission in the active region mainly comes from thermal bremsstrahlung emission. Of course, the gyroresonance emission should also have a contribution to the observed $T_b$, and could be one of the main reasons of the discrepancy between the predicted $T_b$ and the observed one.

\begin{figure}[!ht]
\centering
\includegraphics[width=15cm]{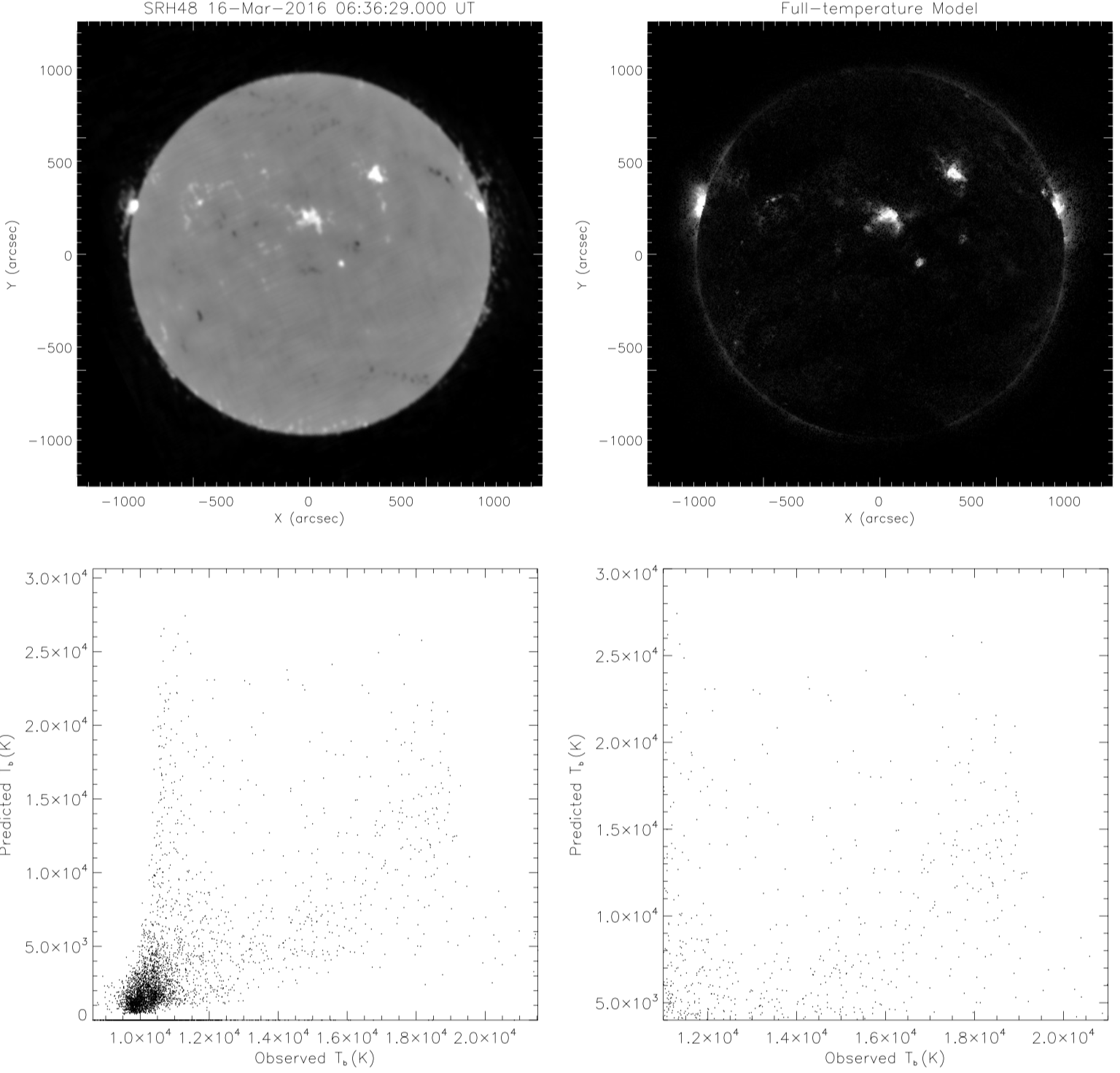}
\caption{\textit{NoRH} full disk radio image at 17 GHz observed on 2016 March 16 (upper left) and the predicted radio image from the full-temperature model at the same frequency and the same time (upper right). Pixel-by-pixel correlation plots between the predicted $T_b$ and the observed one for the full disk (lower left). The lower right panel refers to zooming in of the lower left panel in the high $T_b$ range.}
\label{msfig7}
\end{figure}

In previous studies, \cite{zhang2001reconciling} also found that the predicted radio emission is systematically stronger than the observed value by a factor of 2.0. This work, when using the same two-temperature model as in \cite{zhang2001reconciling}, obtains a much lower predicted value, about 0.36 times the observed one. Such a great difference may be caused by two reasons: the first is that \cite{zhang2001reconciling} did not include the contribution of the plasma with temperatures above $\sim 3$ MK; the second is an underestimation of the iron abundance in their work. The abundance of iron relative to hydrogen for EIT calibration is set to be $3.9\times 10^{-5}$, while that for the AIA calibration is $1.26\times 10^{-4}$ \citep{Meyer, Schmelz2013}. 
We also tentatively study the relationship between the predicted $T_b$ at high frequencies and the observed one. In the upper panels of Figure \ref{msfig7}, we show one full disk image at 17 GHz from NoRH and the predicted $T_b$ image at 17 GHz from the full-temperature model. It is found that for quiescent regions the predicted $T_b$ is much lower than the observed value. For active regions, although the predicted $T_b$ image shows a very similar morphology to the observed one, but their quantitative correlation shown in the scattering plot is really bad as revealed in the lower panels of Figure \ref{msfig7}. Considering the fact that there is little non-thermal particles during non-flaring periods, we think that the radio emission at high frequencies should be probably the gyrosynchrotron emission of the thermal electrons in hot plasma with relatively strong magnetic field. This is the possible reason for the discrepancy between the predicted and observed values of $T_{b}$.

\begin{figure}
\centering
\includegraphics[width=4.9cm]{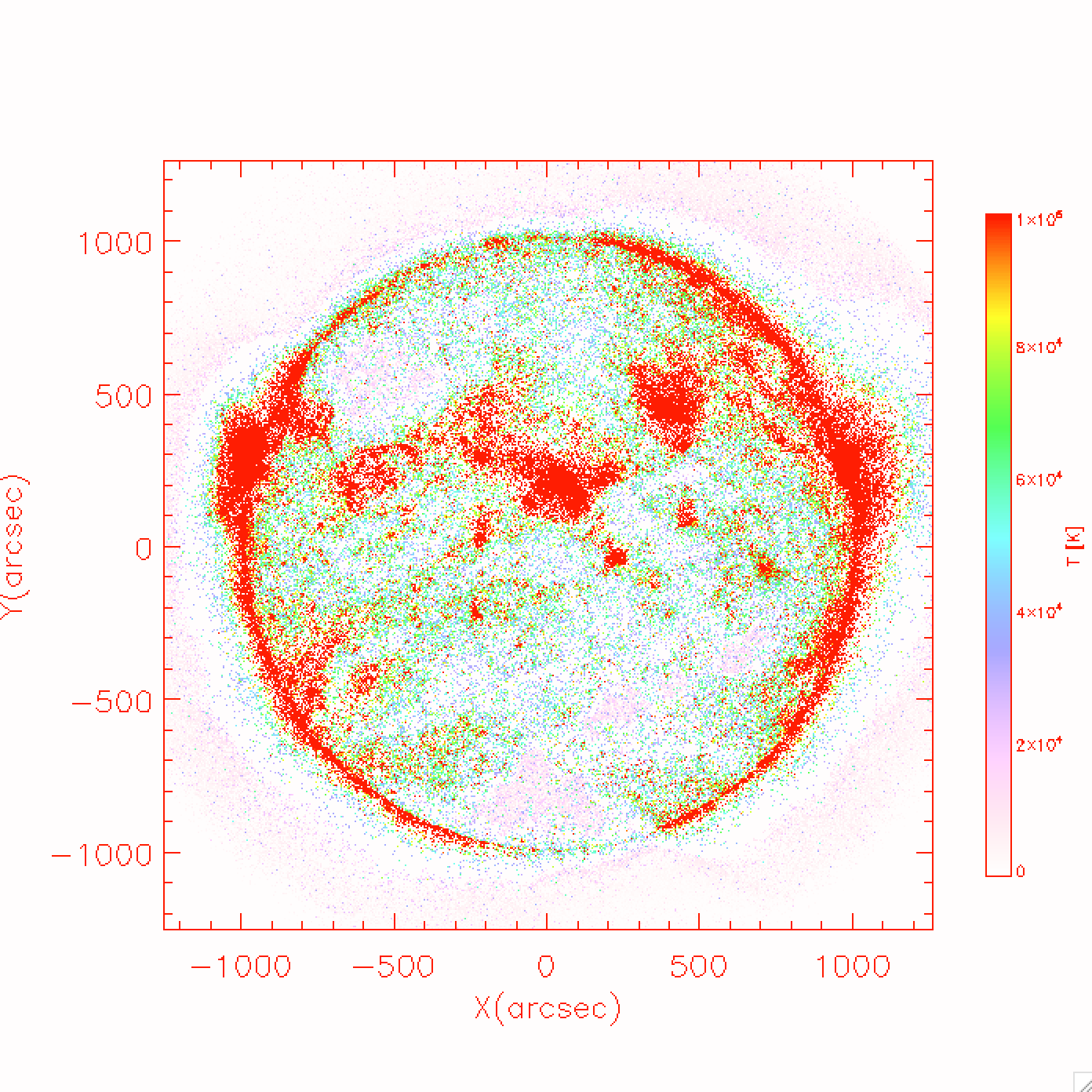}
\includegraphics[width=4.9cm]{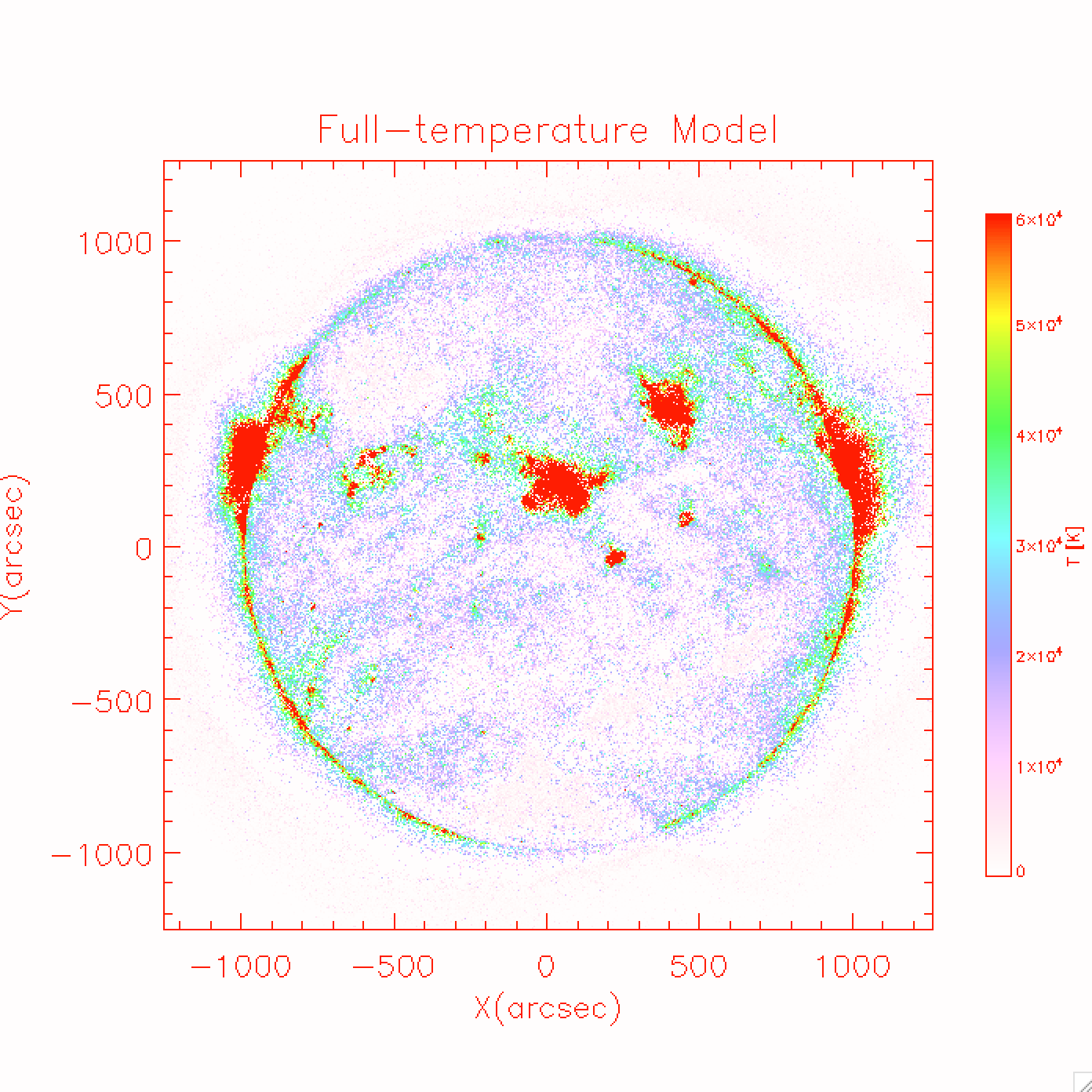}
\includegraphics[width=4.9cm]{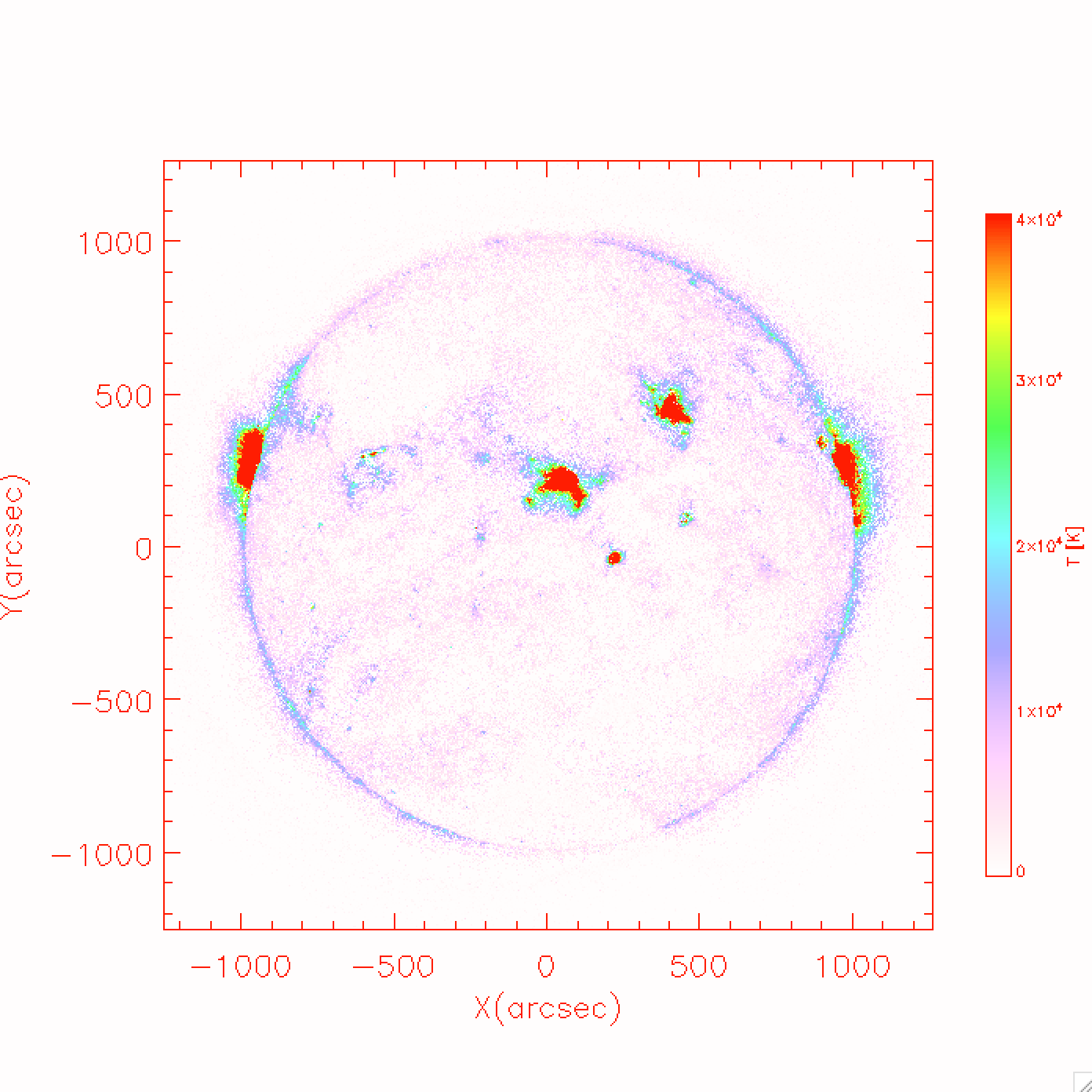}
\caption{Predicted full disk $T_b$ maps at 2.0 GHz, 4.0 GHz, and 8.0 GHz on March 16, 2016.}
\label{msfig8}
\end{figure}

As shown above, there is generally a good linear relationship between the predicted radio $T_b$ at low frequencies, derived from the multi-passband EUV data, and observed one. We also try to predict the $T_b$ maps at other frequencies. In Figure~\ref{msfig8}, we calculate three $T_b$ maps at 2.0 GHz, 4.0 GHz, and 8.0 GHz, respectively, which can be compared with real observations at a wider frequency range such as that from Mingantu Spectral Radioheliograph (MUSER) \citep{Yan2016First} in the near future. We are planning to calculate the $T_b$ maps at four frequencies of 1~GHz, 2~GHz, 4~GHz, and 8~GHz one day per week from 2016 May 16. All data, also including the DEM results, are open for public and can be download at the website of the project \footnote{https://pan.baidu.com/s/1L5kti-oz6z8cXnnZN1ic7w}.

\section{Summary and Discussions}
\label{sect:conclusion}
In this paper, we revisit the relationship between the predicted radio images derived through a two-temperature model and a full-temperature model and the observed radio images during non-flaring times. It is found that the full-temperature model is better to reproduce the observational radio intensity, in particular for active regions. The results confirm that the radio emission of the quiet Sun at low frequencies primarily originates from the thermal plasma in the corona, thus presenting a potentiality for reconciling with MUSER observations in the near future.

However, we also find that the predicted $T_b$ at each pixel not exactly equals to the observed one; the former is usually larger than the latter if using the full-temperature model. Such a discrepancy could be due to the uncertainties in the DEM resolutions derived from the AIA EUV intensities. Note that an inaccurate absolute calibration of radio images could also make the observed flux deviate from the predicted one. Moreover, the predicted radio emission in the full-temperature model is obviously larger than the observed values in the two-temperature model. A possible reason is that the emission measure of the plasma in the full-temperature model is over-estimated. That is to say, an accurate determination of the DEM together with a precise abundance of iron are essential in properly predicting the radio emission. In addition, the gyroresonance emission from thermal electrons can also give rise to a discrepancy between reconstructed $T_b$ from the EUV emission and the observed one.

\begin{acknowledgements}
We are grateful to the referee for constructive comments that helped improve the paper. Z.F.L., S.H.H., X.C., \& M.D.D. are supported by NSFC through grants 11722325, 11733003, 11790303, 11790300 and by Jiangsu NSF through grants BK20170011. X.C. is also supported by ``Dengfeng B" program of Nanjing University.
\end{acknowledgements}

\bibliographystyle{raa}
\bibliography{bibtex}
\label{lastpage}
\end{document}